\documentclass[twocolumn,showpacs,preprintnumbers,amsmath,amssymb]{revtex4}

\usepackage{graphicx}
\usepackage{dcolumn}
\usepackage{bm}

\begin{document} 

\preprint{}
\input{epsf.tex}

\epsfverbosetrue

\title{Excitons and cavity polaritons for ultracold atoms in an optical lattice}
\author{Hashem Zoubi, and Helmut Ritsch}
 
\affiliation{Institut fur Theoretische Physik, Universitat Innsbruck, Technikerstrasse 25, A-6020 Innsbruck, Austria} 
 
\date{29 March, 2007}

\begin{abstract}
We study the resonant electronic excitation dynamics for ultracold atoms trapped in a deep optical lattice prepared in a Mott insulator state. Excitons in these artificial crystals are similar to Frenkel excitons in Noble atom or molecular crystals. They appear when the atomic excited state line width is smaller than the exciton band width generated by dipole-dipole coupling. When the atoms are placed within a cavity the electronic excitations and the quantized cavity mode get coupled. In the collective strong coupling regime excitations form two branches of cavity polaritons with Rabi splitting larger than the atomic and the cavity line width. To demonstrate their properties we calculate the transmission, reflection, and absorption spectra for an incident weak probe field, which show resonances at the polariton frequencies.
\end{abstract}


\maketitle

\section{Introduction}

The quantum phase transition from the superfluid to the Mott insulator phase of cold dilute gas of boson atoms in an optical lattice was widely studied both theoretically \cite{Jaksch,Zwerger}, and experimentally \cite{Bloch}. The optical lattice potential can be formed of counter propagating laser beams, where their interference forms a standing wave with period of $\lambda/2$, half wave length \cite{Morsch}. The geometry and the depth of the optical lattice potential are easily controlled by the laser field. The cold atoms loaded on the optical lattice realize the Bose-Hubbard model \cite{Fisher}, which predicts the phase transition from the superfluid into the Mott insulator phase, by increasing the optical potential depth. The atoms are confined in an array of microscopic trapping potentials due to the optical dipole force, and by changing the relevant parameters, the optical lattice can be filled with few atoms down to one atom per site \cite{Jaksch,Fisher}. Optical lattices allow us to go beyond the weakly interacting regime of Bose gas into strongly correlated systems of condensed matter physics \cite{Zoller}. 

The cold atoms in an optical lattice, in the Mott insulator phase with one atom per site, can be considered as an artificial crystal. Such artificial crystals are similar to Noble atom and molecular crystals, as each atom retains his identity, and negligible overlaps exist between the atom electronic wave functions at different sites. It will be of big importance to check the possibility of the appearance of solid state effects and processes in such artificial crystals. For example in molecular crystals, an electronic excitation localized on a molecule can transfer among the crystal molecules due to electrostatic interactions, i.e. dipole-dipole interactions. Namely, an excited molecule will decay to the ground state and other molecule will be excited, without a charge transfer between the molecules. In using the crystal symmetry, such energy transfer can be represented in the quasi-momentum space by a wave that propagates in the crystal. We obtain quasi-particles with an effective mass and a finite life time, which are called excitons \cite{Agranovich,Zoubi}. If the molecular crystal is located within a cavity, the excitons and the cavity photons are coupled. In the strong coupling regime, where the exciton and photon line-widths are smaller than the coupling strength, the excitons and photons are coherently mixed to form new system quasi-particles which are called cavity-polaritons \cite{Kavokin,Zoubi}. We get two polariton branches which are separated by the Rabi splitting energy. Coherent interactions of laser pulses in a resonant optically dense medium of two-level atoms was presented in \cite{Mekhov}.

In the present paper we investigate such excitons and polaritons. We consider two-level atoms within a cavity, with transition energies close to resonance with a cavity mode. The optical lattice potentials are chosen to give light shifts in-phase for the ground and excited states, with minimums at the same positions. In multilevel atoms and for a specific laser frequency one can achieve an excited level with light shifts in-phase with the ground state ones, and there even exists a magic frequency where we get equal shifts as in the ground state \cite{Zoller,Katori}. In all the discussions the trapping fields are considered classical and far from resonance with any of the atom electronic transitions. Furthermore, we limit the treatment to the Mott insulator phase with one atom per site \cite{Chen}. We examine the possibility of the appearance of excitons in such a system, by comparing between the energy transfer coupling parameter induced by the dipole-dipole resonance interaction, and the excited state line width. In a cavity the view is different, now the excitons are strongly coupled to the cavity photons by the dipole interaction. Hence, coherent polariton states can be observed which are superpositions of excitons and cavity photons. In paper \cite{Ritsch}, the cavity photons are also treated quantum mechanically, but their energies are far to resonance with any of the atom transitions, and they are used to produce the optical lattice. Moreover, we get the linear optical spectra of the cavity for an incident external field.

The paper is organized as follows. In section 2 we discuss excited and ground state cold atoms loaded on an optical lattice. The excitons in such a system are introduced in section 3. Section 4 describes polaritons in a system of optical lattice cold atoms within a cavity. The transmission, reflection, and absorption spectra are given in section 5. The conclusions are presented in section 6. In the appendix the linear optical spectra are derived using the input-output formalism.

\section{Excited and ground state cold-atoms in an optical-lattice}

We consider neutral cold two-level atoms where the cavity mode is close to resonance with only a single excited state. The cold atoms are trapped on $2D$ standing laser waves far to resonance with any of the atomic excitations, and result in shifts in the ground and excited states. Such shifts correspond to an optical lattice potential $V_g({\bf x})$ for ground state atoms, and $V_e({\bf x})$ for excited state atoms, where ${\bf x}$ is the atom location. The potentials are periodic with a periodicity of, $\lambda/2$, half laser wave length. The potentials are in general out of phase, and for two-level atoms they have $\pi/2$ phase shift, where the position of the minimum of one potential appears at the maximum of the other. Even though, we assume the two optical lattice potentials to be in-phase, with minimums located at the same position. Such assumption is possible as for multi-level atoms, i.e. Alkali or Alkaline atoms, one get a number of optical lattices, one for each level. For a specific laser magic frequency and polarization exists an excited state optical lattice potential is equal to the ground state one \cite{Zoller,Katori}. The two dimensional optical lattice potentials are $V_s({\bf x})=\sum_iV_s\ \cos^2(kx_i)$, with $(x_i=x,y)$, and $(s=g,e)$, where $k=2\pi/\lambda$, and $V_s$ is the $s$ potential depth, see figure (1). Furthermore, the ground and excited state atoms are located in the first Bloch band of the ground and excited optical lattice potentials, which is the case for deep optical lattices.

\begin{figure}[h!]
\centerline{\epsfxsize=8.0cm \epsfbox{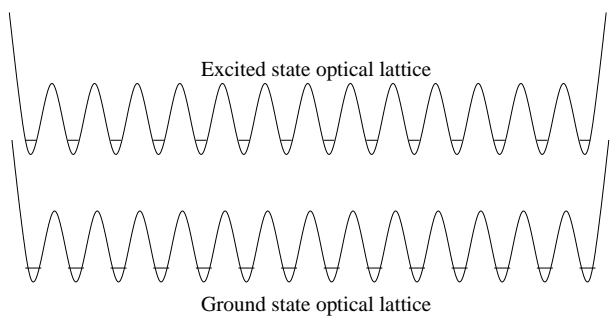}}
\caption{Schematic plot of the ground and excited state optical potentials.}
\end{figure}

The ground and excited state atoms are considered as two kinds of bosons, which are represented by the Bose-Hubbard model of two components of bosons \cite{Chen}, that yields a rich phase diagram with two main phases, the Mott insulator and the superfluid phases. In the present paper we consider only the Mott insulator phase with one atom per site, which was realized experimentally \cite{Bloch}.

The internal two-state Hamiltonian of a free atom is $H_A=\hbar\omega_g'|g'\rangle\langle g'|+\hbar\omega_e'|e'\rangle\langle e'|$, where $|g'\rangle$ and $|e'\rangle$ are the ground and excited states with energies $\hbar\omega_g'$ and $\hbar\omega_e'$, respectively. Now the atoms are loaded on an optical lattice, and therefore the atom internal states become a function of the atom position on the optical lattice. In the Mott insulator phase with one atom per site, and for the first Bloch band, the Hamiltonian of the atoms reads
\begin{equation}
H_A=\sum_i\left\{\hbar\omega_g|g_i\rangle\langle g_i|+\hbar\omega_e|e_i\rangle\langle e_i|\right\}.
\end{equation}
The summation is over the optical lattice sites. The optical lattice in this case results, first in energy shifts of the atom internal states, where we have now the energies $\hbar\omega_g$ and $\hbar\omega_e$ in place of the free atom ones, and the atomic energies are the same at each site. At the laser magic frequency the atom frequency transition $\omega_e-\omega_f$ equals to the free atom transition $\omega_e'-\omega_f'$. Second, the optical lattice limits the atom positions to be confined on a lattice with lattice constant $a=\lambda/2$, which is indicated by the index $i$ in the above Hamiltonian.

Each atom retains his identity. Namely, as the average distance between the atoms is of the order of the lattice constant $a$, no overlaps of the electronic wave functions exist between atoms at different sites. This fact makes such artificial atomic lattices similar to Nobel atom or Molecular crystals. Here the crystal forms of external laser fields which can be controlled, e.g. it is possible to control the lattice constant and the number of atoms per site. Even though, the electronic excitations can transfer between atoms at different sites due to electrostatic interactions between the atoms, i.e. dipole-dipole interactions, as an excited atom has a transition electric dipole. An excited atom at site $i$ jumps to the ground sate and another atom at site $j$ is excited, no charge transfer exists between the atoms of such interactions, see figure (2) for $1D$ case. The resonance energy transfer can be induced by dipole-dipole interactions, or higher order multi-pole interactions, with the energy transfer parameter $J_{ij}$. The energy transfer Hamiltonian between atoms is given by
\begin{equation}
H_T=\sum_{i,j}\hbar J_{ij}\ S_i^{\dagger}S_j,
\end{equation}
where $S_i^{\dagger}=|e_i\rangle\langle g_i|$ and $S_i=|g_i\rangle\langle e_i|$, are creation and annihilation operators of an excitation in the atom at site $i$.

\begin{figure}[h!]
\centerline{\epsfxsize=8.0cm \epsfbox{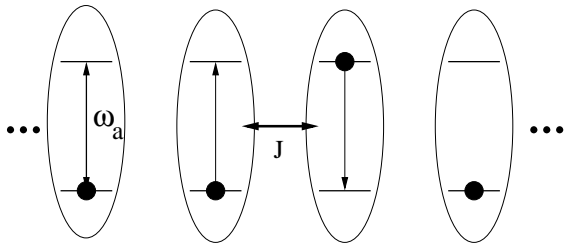}}
\caption{Energy transfer in $1D$ atoms chain.}
\end{figure}

The atomic Hamiltonian, in terms of internal excitation creation and annihilation operators, is given by
\begin{equation}
H_a=\sum_i\hbar\omega_a\ S_i^{\dagger}S_i+\sum_{i,j}\hbar J_{ij}\ S_i^{\dagger}S_j,
\end{equation}
where $\omega_a=\omega_e-\omega_g$. The Hamiltonian includes only terms that conserve the excitation number.

\section{Excitons in optical lattice cold atoms}

The electronic excitations can transfer among the atoms in the optical lattice with energy transfer parameter $J_{ij}$, between two atoms at sites $i$ and $j$. The coupling parameter is a function of the distance between the atoms, $J_{ij}=J(|{\bf n}_i-{\bf n}_j|)$, where ${\bf n}_i$ is the atom $i$ location in the lattice. The excitation can be at any atom with the same probability, and in using the lattice symmetry the excitation is represented in the quasi-momentum space by a wave propagates in the lattice with wave vector ${\bf k}$. Such a quasi-particle is called Frenkel-exciton in molecular crystals, e.g. in Nobel atom or organic crystals \cite{Agranovich,Zoubi}. Here we study Frenkel like-excitons in optical lattice cold atoms, and we give the condition for obtaining such quasi-particles in such a system.

The operators $S_i$ are of two-level atom which are as for a system of spin-half lattice. They obey Bose commutation relation between different sites, namely $[S_i,S_j^{\dagger}]=0$, and Fermi anti-commutation relation on the same site, namely $\{S_i,S_i^{\dagger}\}=1$, which forbids two excitations from being localized on the same atom. At low excitation density we neglect the possibility of getting two excitations on the same atom, and we can assume the excitations to behave as Bosons \cite{ZoubiA}. The bosonization procedure is done by making the replacement $S_i\rightarrow B_i$, where $[B_i,B_j^{\dagger}]=\delta_{ij}$. Namely we neglect saturation effects which give rise to nonlinear processes. We treat the single excitation case, which is given by the bosonic Hamiltonian
\begin{equation}
H_a=\sum_i\hbar\omega_a\ B_i^{\dagger}B_i+\sum_{i,j}\hbar J_{ij}\ B_i^{\dagger}B_j.
\end{equation}
The Hamiltonian can be easily diagonalize by the transformation into the quasi-momentum space, in using
\begin{equation}\label{EXCITON}
B_i=\frac{1}{\sqrt{N}}\sum_{\bf k}B_{\bf k}\ e^{i{\bf k}\cdot{\bf n}_i},
\end{equation}
where $N$ is the number of lattice sites. The exciton Hamiltonian reads
\begin{equation}
H_a=\sum_{\bf k}\hbar\omega_a({\bf k})\ B_{\bf k}^{\dagger}B_{\bf k},
\end{equation}
with the exciton dispersion
\begin{equation}
\omega_a({\bf k})=\omega_a+\sum_{\bf L}J({\bf L})\ e^{-i{\bf k}\cdot{\bf L}},
\end{equation}
where we defined the distance between atoms by ${\bf L}={\bf n}_i-{\bf n}_j$. 

Our system is $2D$ square with a cubic symmetry. The in-plane wave vector is defined by ${\bf k}=(k_x,k_y)=(n_x,n_y)\times 2\pi/\sqrt{S}$, where the system area is $S=Na^2$, and we have $(n_x,n_y=0,\pm 1,\cdots,\pm \sqrt{N}/2)$. If we assume energy transfer only between the first two nearest neighbor sites with coupling parameters $J(a)=-J_1$ and $J(\sqrt{2}a)=-J_2$, which are negative and give attractive interactions, hence the exciton dispersion reads
\begin{eqnarray}
\omega_a({\bf k})&=&\omega_a-2J_1\ (\cos k_xa+\cos k_ya) \nonumber \\
&-&4J_2\ \cos (\sqrt{2}ak_x)\ \cos (\sqrt{2}ak_y).
\end{eqnarray}
The energy transfer in a symmetric lattice results in an energy band, in place of a discrete level for independent atoms. In considering only the first nearest neighbor interaction, the band width is $4\hbar J_1$. In fact the dipole-dipole interaction is of long-range interaction, and hence one needs to go beyond the nearest neighbor interactions. But here as the lattice constant is of the order of the atom transition wave length, we limited the discussion for nearest neighbor interactions.

In the long wave length limit, that is $ka\ll 1$ with $k=|{\bf k}|$, which is the regime of importance for the excitation-photon coupling as we will see in the next section, we get the parabolic dispersion
 \begin{equation}
\omega_a(k)=\omega_a-4(J_1+J_2)+\frac{\hbar k^2}{2m_{eff}},
\end{equation}
where we defined the exciton effective mass by $m_{eff}=\hbar/[2a^2(J_1+4J_2)]$. Note that at zero wave vector, $k=0$, a significant atomic frequency shift of $4(J_1+J_2)$, relative to the free atom frequency, is obtained and which can be easily observed. Also such a shift has influence on optical lattice clocks \cite{Lukin}. 

The exciton coherent states defined in Eq.(\ref{EXCITON}) compete with the excited state life time $\tau_a$. In order to observe exciton effects in optical lattice cold atoms, the exciton life time need to be longer than the energy transfer time $1/J_1$ between nearest neighbor sites; or the excitation line width, $\hbar \gamma_a=\hbar /\tau_a$, need to be smaller than the exciton band width of the nearest neighbor interaction, namely $\hbar \gamma_a<4\hbar J_1$. For example, in free $^{85}Rb$ for the transition $5^2S_{1/2}-5^2P_{3/2}$ we have $\hbar\omega_a=1.56\ eV$, with line width of $\hbar\gamma_a=2.5\times 10^{-8}\ eV$; and for free $^{23}Na$ for the transition $3^2S_{1/2}-3^2P_{3/2}$ we have $\hbar\omega_a=2.1\ eV$, with $\hbar\gamma_a=4\times 10^{-8}\ eV$. The transfer energies $\hbar J_1$ and $\hbar J_2$ are calculated from the dipole-dipole interaction between two atom transition dipoles $\vec{\mu}^1$ and $\vec{\mu}^2$ which are separated by the distance vector $\vec{R}=R\hat{R}$, which is given by \cite{Craig}
\begin{eqnarray}
\hbar J(R)&=&\sum_{i,j}\frac{\mu_i^1\mu_j^2}{4\pi\epsilon_0R^3}\left\{\left(\delta_{ij}-3\hat{R}_i\hat{R}_j\right)\left(\cos lR+lR\ \sin lR\right)\right. \nonumber \\
&-&\left.\left(\delta_{ij}-\hat{R}_i\hat{R}_j\right)\ l^2R^2\ \cos lR\right\},
\end{eqnarray}
here $(i,j=x,y,z)$, with $\omega_a=cl$. We assume $\vec{R}=R\hat{R}_z$, and for the case of two equal dipoles in the $z$ direction, where $\vec{\mu}^1=\vec{\mu}^2=\mu\hat{R}_z$, we get the coupling
\begin{equation}
\hbar J(R)=-\frac{\mu^2}{2\pi\epsilon_0R^3}\left(\cos lR+lR\ \sin lR\right).
\end{equation}
The distance between two atoms, for the nearest neighbor coupling, equals the lattice constant, that is $R=a$. For atoms with $\hbar\omega_a=2\ eV$, dipole of $\mu=2\ e\AA$, and optical lattice of $a=1000\ \AA$, we get $\hbar J_1=10^{-7}\ eV$, which is one order of magnitude larger than the free atom line width, and excitons can be observed in this case. For the next nearest neighbor interaction we get $\hbar J_2=3\times 10^{-8}\ eV$, which is of the order of the free atom line width. The higher nearest neighbor transfer interaction terms are much smaller than the free atom line width, and they are not relevant for the exciton formation, and hence can be neglected here.

In the next section we show how in using a cavity one can observe better coherent effects involving cavity photons, where the excitation transfer is mediated by the cavity photons, in the strong coupling regime, where the excitation-photon coupling is larger than the atomic and cavity-photon line widths.

\section{Cavity polaritons in a cold atom crystal}

Let us now consider such a quantum gas of ultracold atoms enclosed in the middle of optical resonator built of two parallel mirrors, which we consider as perfect for the moment (see figure (3)). Note that in contrast to most cavity QED configurations, where one considers single modes of spherical mirror resonators with Gaussian transverse geometry, our cavity is adapted to the translational invariance of the lattice supporting plane wave field modes. 
 
\begin{figure}[h!]
\centerline{\epsfxsize=8.0cm \epsfbox{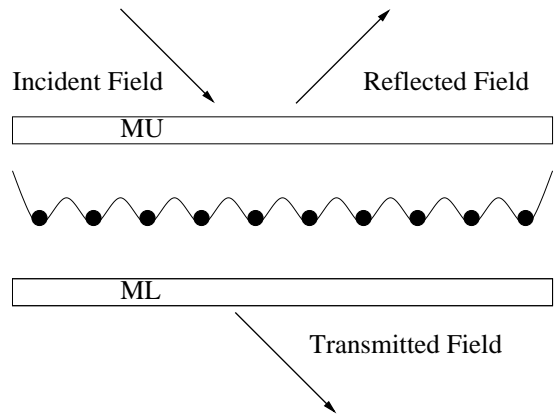}}
\caption{Optical lattice cold atoms within a cavity.}
\end{figure}

As the electromagnetic field is not confined in the cavity plane with in-plane wave vector ${\bf q}$, the modes have discrete wave vector $k_z=m\pi/L$, where $m=1,2,\cdots$ only in the perpendicular $z$ direction. As the cavity distance in $z$ direction is small we can restrict our considerations to only the $m$-mode close to resonance with the atomic electronic excitation. As the optical lattice is located in the middle between the cavity mirrors the active mode is taken to be one of the odd modes to ensure strong coupling. For each in-plane wave vector exist two possible polarizations, $TE$ and $TM$ modes. Here we consider only isotropic materials so that we can concentrate in a fixed cavity photon polarization adapted to the atomic excitations. The cavity Hamiltonian thus is given by \cite{Haroche}
\begin{equation}
H_c=\sum_{\bf q}\hbar\omega_c(q)\ a^{\dagger}_{\bf q}a_{\bf q},
\end{equation}
where $a^{\dagger}_{\bf q},\ a_{\bf q}$ are the creation and annihilation boson operators of a cavity photon with in-plane wave vector ${\bf q}$. The corresponding cavity photon dispersion thus reads
\begin{equation}
\omega_c(q)=\frac{c}{\sqrt{\epsilon}}\sqrt{q^2+\left(\frac{m\pi}{L}\right)^2},
\end{equation}
 where $L$ is the distance between the cavity mirrors, and $\epsilon \approx 1$ is the cavity medium dielectric constant. Note the following discussions also hold for the case of closed cavities with discrete and completely confined modes. Single mode operation corresponds to limiting ourselves to the case of zero in-plane wave vector, $k=0$, and with a fixed detuning between the atom transition and the cavity mode.

The atom electronic transitions  and thus the excitons are coupled to the cavity modes by the dipole interaction $V=-\hat{\mu}\cdot{\bf E}$, where $\hat{\mu}=\vec{\mu}\sum_i(S_i^{\dagger}+S_i)$ is the dipole operator and ${\bf E}$ is the cavity field operator. Assuming equal dipole moments for all the atoms and the Mott insulator phase with one atom per site, the interaction Hamiltonian in the rotating wave approximation reads:
\begin{equation}
V=\sum_{{\bf k}i}\left\{\hbar f_{{\bf k}i}\ S_i^{\dagger}a_{\bf k}+f^{\ast}_{{\bf k}i}\ a_{\bf k}^{\dagger}S_i\right\},
\end{equation}
where the coupling parameter is
\begin{equation}
\hbar f_{{\bf k}i}=-i\sqrt{\frac{\hbar\omega_{c}(k)}{2LS\epsilon_0}}\ (\vec{\mu}\cdot\hat{\epsilon}_{\bf k})e^{i{\bf k}\cdot{\bf n}_i},
\end{equation}
$S$ is the mirror area, and $\hat{\epsilon}_{\bf k}$ is the photon polarization unit vector. In using the previous bosonization procedure, and in applying the transformation (\ref{EXCITON}), we get
\begin{equation}
V=\sum_{\bf k}\hbar \left(f_k\ B^{\dagger}_{\bf k}a_{\bf k}+f^{\ast}_k\ a^{\dagger}_{\bf k}B_{\bf k}\right),
\end{equation}
with
\begin{equation}
\hbar f_k=-i\sqrt{\frac{\hbar\omega_{c}(k)N\mu^2}{2LS\epsilon_0}}.
\end{equation}

Here we assumed an isotropic atomic ground state so that for each direction of ${\bf k}$ we have the same dipole moment $\mu$. The atom position dependence disappears due to the lattice translational symmetry, and the lattice symmetry ensures
that only coupling between excitons and photons with the same in-plane wave vector occurs, i.e. we have quasi-momentum conservation. The coupled exciton-photon Hamiltonian is written as
\begin{eqnarray}
H&=&\sum_{\bf k}\hbar\left\{\omega_a(k)\ B_{\bf k}^{\dagger}B_{\bf k}+\omega_c(k)\ a^{\dagger}_{\bf k}a_{\bf k}\right. \nonumber \\
&+&\left.f_k\ B^{\dagger}_{\bf k}a_{\bf k}+f^{\ast}_k\ a^{\dagger}_{\bf k}B_{\bf k}\right\}.
\end{eqnarray}

Due to exciton-photon coupling the exciton and the photon coherently mix and it makes sense to transform to
a diagonal basis with respect to this coupling, namely the polariton basis \cite{Kavokin,Zoubi}. The maximal coupling appears at small wave vectors as the exciton band width is narrow relative to the photon band and the exciton effective mass thus is much larger than the cavity photon effective mass. Therefore the exciton dispersion around the exciton-photon coupling can be neglected, and one can use $\omega_a(k)\approx\omega_a-4(J_1+J_2)$.

The above Hamiltonian can be easily diagonalized to give
\begin{equation}
H=\sum_{{\bf k}r}\hbar\Omega_r(k)\ A^{\dagger}_{{\bf k}r}A_{{\bf k}r},
\end{equation}
which exhibits two diagonal polariton branches with different dispersions
\begin{eqnarray}
\Omega_{\pm}(k)=\frac{\omega_{c}(k)+\omega_a}{2}\pm\Delta_k,
\end{eqnarray}
where $\Delta_k=\sqrt{\delta_k^2+|f_k|^2}$ and we defined the exciton-photon detuning by $\delta_k=(\omega_{c}(k)-\omega_a)/2$. The splitting between the two polariton branches at wave vector ${\bf k}$ is $2\Delta_k$ and the splitting at the exciton-photon intersection point, where $\delta_k=0$, is $2|f_k|$ which exactly corresponds to the vacuum Rabi splitting.

The polaritons thus are coherent superpositions of excitons and photons with operators
\begin{eqnarray} \label{TRANSFOR}
A_{{\bf k}\pm}=X_{k}^{\pm}B_{\bf k}+Y_{k}^{\pm}a_{{\bf k}},
\end{eqnarray}
where the exciton and photon amplitudes are
\begin{eqnarray}
X_{k}^{\pm}=\pm\sqrt{\frac{\Delta_k\mp\delta_k}{2\Delta_k}}\ ,\ Y_{k}^{\pm}=\frac{f_{k}}{\sqrt{2\Delta_k(\Delta_k\mp\delta_k)}}.
\end{eqnarray}
At the exciton-photon intersection point the polaritons are half exciton and half photon, that is $|X_{k}^{\pm}|^2=|Y_{k}^{\pm}|^2=1/2$. At large wave vectors the lower branch becomes excitonic, that is $|X_{k}^{-}|^2\approx1,\ |Y_{k}^{-}|^2\approx0$, and the upper branch becomes photonic, that is $|X_{k}^{+}|^2\approx0,\ |Y_{k}^{+}|^2\approx1$.

In figure (4) we illustrate these results and plot the upper and lower polariton branches beside the cavity photon and exciton dispersions using the following typical numbers. The exciton energy around small in-plane wave vector is taken to be $\hbar\omega_a-4(J_1+J_2)=2\ eV$. The distance between the cavity mirrors is $L/m=3100\ \AA$, which is chosen to give zero detuning between the exciton and the cavity photon dispersions at zero in-plane wave vector. For example for $m=3$, we have $L\approx 1\ \mu m$. The transition dipole is $\mu=2\ e\AA$, and the optical lattice constant is $a=2000\ \AA$. The exciton-photon coupling parameter, by using $S=Na^2$, is given by $|f|=\sqrt{(\omega_c(0)\mu^2)/(2\hbar La^2\epsilon_0)}$, with the value of $|f|=4\times 10^{11}\ Hz$, where we neglected the $k$ dependence for small in-plane wave vectors. At zero in-plane wave vector the Rabi splitting frequency of $8\times 10^{11}\ Hz$ is clear. At large wave vectors the upper polariton branch tends to the cavity photon dispersion, and the lower branch tends to the exciton dispersion. 

In figure (5) we plot the exciton and photon weights in the lower polariton branch, $|X^-|^2$ and $|Y^-|^2$. At zero in-plane wave vector the polariton is half exciton and half photon. For large in-plane wave vectors the polariton becomes much more excitonic than photonic. In figure (6) we plot the exciton and photon weights in the upper polariton branch, $|X^+|^2$ and $|Y^+|^2$. At zero in-plane wave vector the polariton is half exciton and half photon. For large in-plane wave vectors the polariton becomes much more photonic than excitonic.

\begin{figure}[h!]
\centerline{\epsfxsize=8.0cm \epsfbox{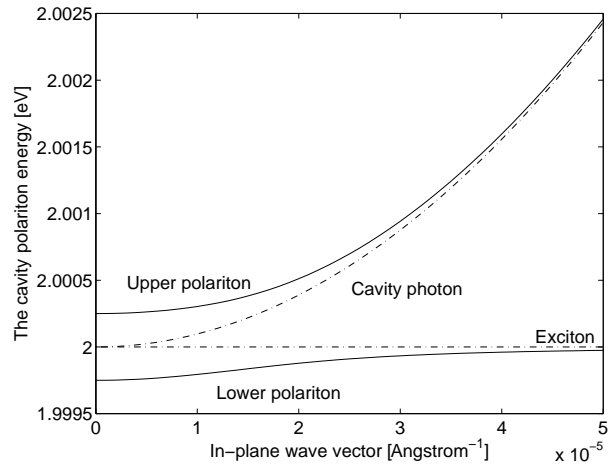}}
\caption{The upper and the lower polariton branches vs. in-plane wave vector $k$. The dashed-dotted line is for the exciton dispersion, and the dashed-dotted parabola is for the cavity photon dispersion.}
\end{figure}

\begin{figure}[h!]
\centerline{\epsfxsize=8.0cm \epsfbox{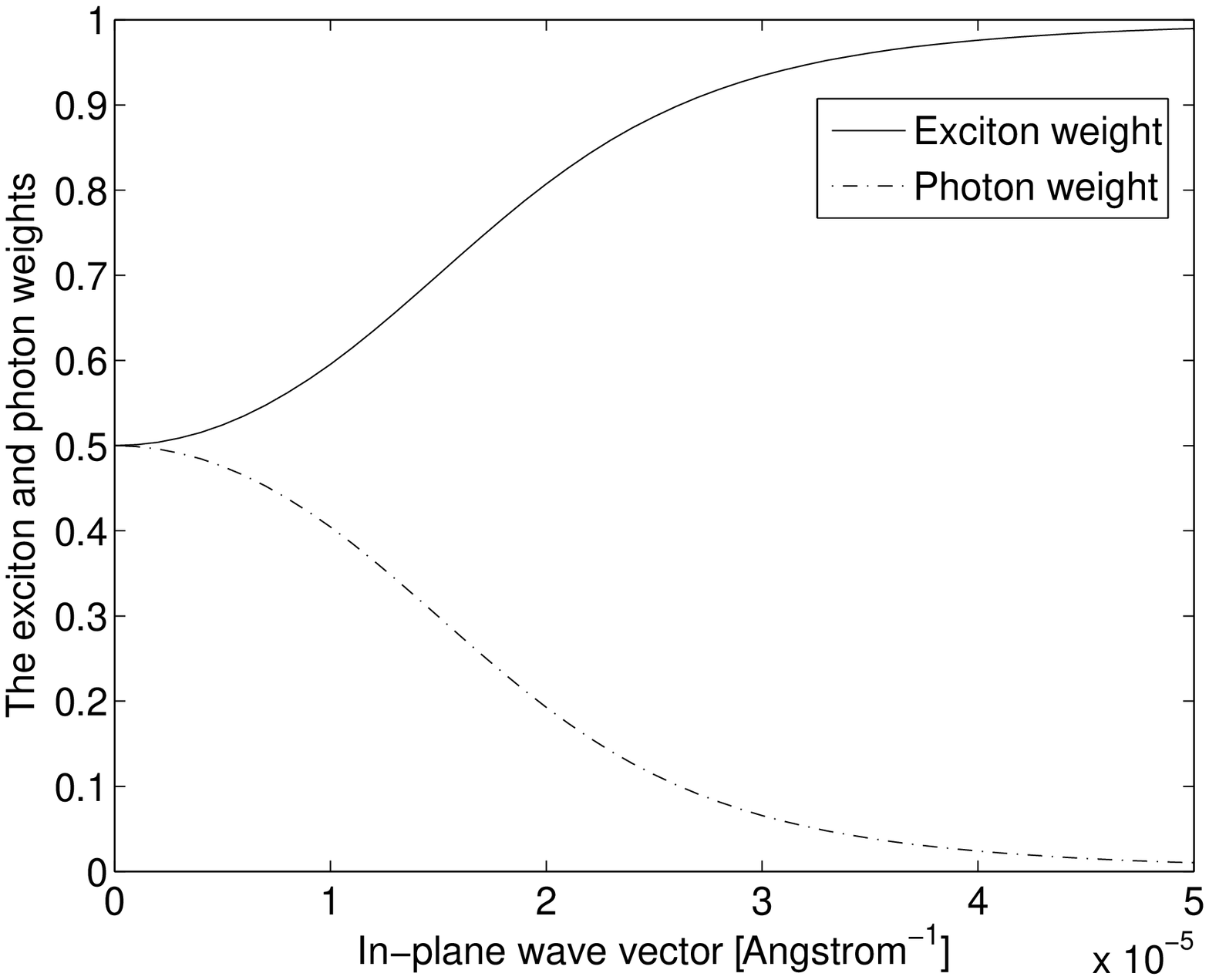}}
\caption{The exciton and photon weights vs. in-plane wave vector $k$, in the lower polariton branch.}
\end{figure}

\begin{figure}[h!]
\centerline{\epsfxsize=8.0cm \epsfbox{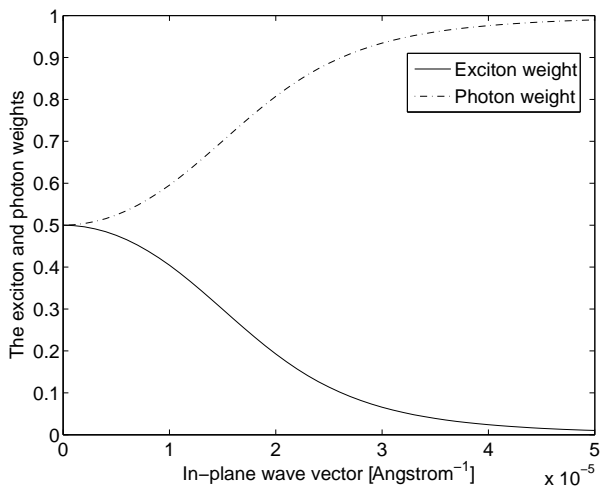}}
\caption{The exciton and photon weights vs. in-plane wave vector $k$, in the upper polariton branch.}
\end{figure}

Let us now discuss some dynamical consequences of this coupling and discuss the possibility to generate Bose-Einstein Condensation (BEC) of polaritons analogous to recent setups in semiconductors \cite{Kasprzak} for Wannier-Mott cavity polaritons. This process could be started by exciting lower branch polaritons at high energies. In turn the polaritons relax along the lower branch toward the minimum energy at $k=0$. If one can achieve sufficient accumulation of polaritons at $k=0$ a phase transition into the BEC can be expected. The phase transition is stimulated by the nonlinear polariton-polariton interactions originated of the atom saturations. The relaxation towards $k=0$ is induced mainly by polariton-polariton interactions and to some extend by the emission of on-site vibrational quanta, which corresponds to localized atomic excitation to a higher Bloch band. As we neglect direct atomic tunneling such on-site excitations are dispersion-less, so that no bottle-neck effect exists in the present system. In semiconductor microcavities the main relaxation mechanism is due to the scattering of polaritons by acoustic-phonons and as the lower polariton branch slope around the exciton-photon intersection point is larger than the sound velocity in semiconductor crystals, the polaritons are stuck near this momentum in a so called bottle-neck effect \cite{Kavokin}. This effect is considered as one of the main obstacles for achieving a BEC of polaritons in semiconductor microcavities \cite{Kasprzak}. Here we are centrally limited by the short life time of the atomic excited states and the cavity photons. The atomic decay time could in principle be enlarged within a cavity to get an effective polariton life time longer than the polariton relaxation. One alternative would be the use of long lived (metastable) atomic states and very high finesse mirrors. 

\section{Polartion analysis via linear cavity transmission spectra}

Up to now we considered an ideal cavity with perfect mirrors. But in order to observe the system eigenmodes, we need to couple the internal cavity modes to the external world. We do this in using the standard input output formalism or quasi-mode approach \cite{Gardiner}, which is well proved for high-Q cavities. In this approach the internal cavity field modes and the external free radiation field are quantized separately and weakly coupled via the cavity mirrors. This leads to the cavity field damping (photon loss) but also allows to consistently calculate the relation between the cavity input and output fields. To include atomic spontaneous emission as an energy loss term we add an effective exciton damping phenomenologically. This finally allows to calculate the transmission $T$, reflection $R$, and absorption $A$ spectra for a given incident field as shown in figure (3). The details of the calculations are presented in the appendix, which yield
\begin{eqnarray} \label{SPECTRA}
T(\omega)&=&\frac{\gamma_U\gamma_L\ |\Lambda(\omega)|^2}{|D(\omega)|^2}, \nonumber \\
R(\omega)&=&\frac{1-i\bar{\gamma}[\Lambda(\omega)-\Lambda^{\ast}(\omega)]+\bar{\gamma}^2|\Lambda(\omega)|^2}{|D(\omega)|^2}, \nonumber \\
A(\omega)&=&\frac{i\gamma_U[\Lambda(\omega)-\Lambda^{\ast}(\omega)]}{|D(\omega)|^2},
\end{eqnarray}
where we defined $\Lambda(\omega)=\sum_r|Y^r|^2/(\omega-\bar{\Omega}_r)$, $\bar{\gamma}=(\gamma_U-\gamma_L)/2$, $|D(\omega)|^2=|1+i\gamma\ \Lambda(\omega)|^2$, and the other parameters are defined in the appendix.

Now we present plots of the transmission, reflection, and absorption spectra. We consider the system given in the previous section and we plot the spectra for different wave vectors. Here we choose the following numbers for the damping rates. The exciton damping rate is $\Gamma_{ex}=1.5\times 10^{9}\ Hz$, the upper and lower mirror damping rates are $\gamma_U=\gamma_L=7.5\times10^{10}\ Hz$. In figure (7) we plot the transmission, in figure (8) the reflection, and in figure (9) the absorption. The peaks of the transmission and the absorption, and the dips of the reflection correspond to the two polariton branches, which are the real eigenmodes of the system. From the plots we conclude that large absorption is obtained for polaritons which are much more excitonic than photonic. High transmission peaks and deep reflection dips are obtained for polaritons which are much more photonic than excitonic. At zero wave vector, as the polariton is half exciton and half photon, the upper and the lower spectra are identical. As the wave vector increase the upper branch becomes more photonic than excitonic, and the opposite for the lower one. Hence, the line-width of the upper branch spectrum becomes wider, which equals the cavity line-width, and the lower one becomes narrower, which equals the atom line-width. As the absorption is of the atomic excitations, at large wave vectors absorption only of the lower branch exists.

\begin{figure}[h!]
\centerline{\epsfxsize=8.0cm \epsfbox{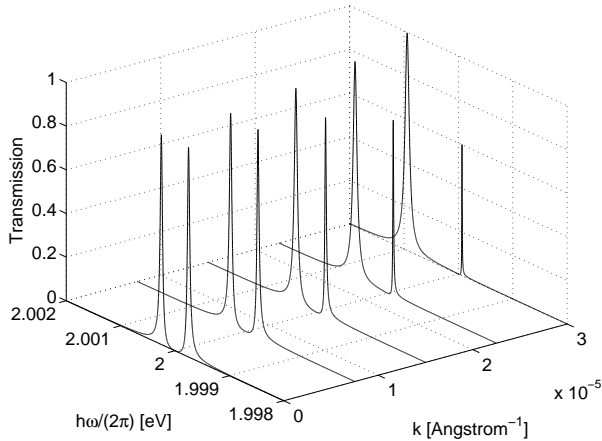}}
\caption{The transmission spectrum for different in-plane wave vectors (the used parameters appear in the text).}
\end{figure}

\begin{figure}[h!]
\centerline{\epsfxsize=8.0cm \epsfbox{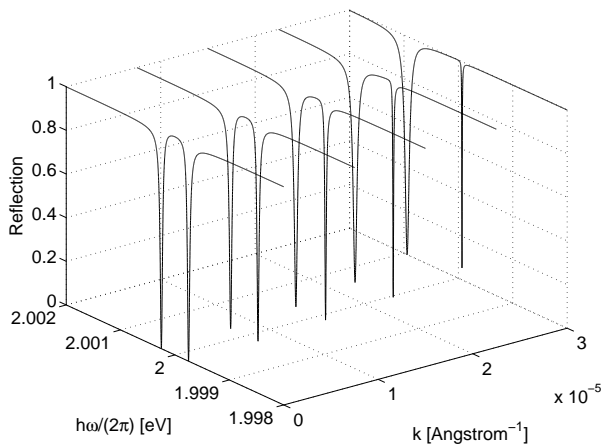}}
\caption{The reflection spectrum for different in-plane wave vectors (the used parameters appear in the text).}
\end{figure}

\begin{figure}[h!]
\centerline{\epsfxsize=8.0cm \epsfbox{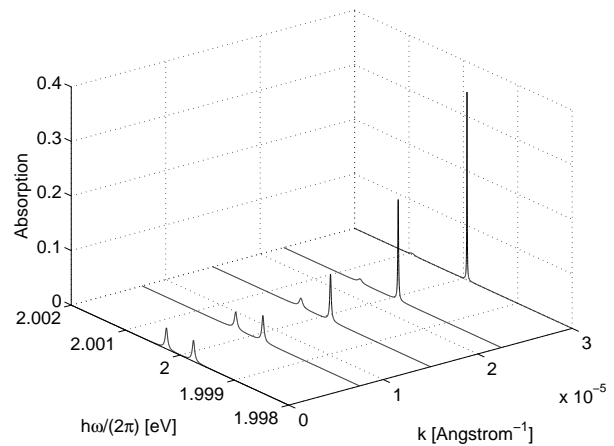}}
\caption{The absorption spectrum for different in-plane wave vectors (the used parameters appear in the text).}
\end{figure}

\section{Conclusions}

We presented a theory of excitons in ultracold atoms in an optical lattice. When the system is placed in a cavity, the eigenstates can be well described in terms of cavity polaritons. The atoms are taken to be two-level systems, where we chose two-levels close to resonance with a cavity mode, and their optical lattice potentials have equal light shifts. We limited ourselves to the Mott insulator phase with one atom per site. An excitation can transfer among the lattice atoms due to dipole-dipole interactions. The lattice symmetry allow us to present an excitation transfer by a wave that propagates in the lattice. These coherent states are Frenkel like-excitons, as in molecular crystals. We showed that excitons can be observed in optical lattices only if the atom line width is smaller than the exciton band width, which is achieved for optical lattices with a small lattice constant. After that, the system is fabricated between cavity mirrors. The excitons are coupled to the cavity photons by the electric dipole interaction. In the strong coupling regime the system eigenmodes are polaritons, which are coherent superpositions of excitons and photons. The polaritons can be observed directly by linear optical spectra. The transmission, reflection, and absorption spectra for an incident external field show resonances at the polariton energies. The system presented here will have big applications in optoelectronic devices and quantum information, and can open new directions in optical lattice systems, both theoretically and experimentally. Also the polaritons can be used as a strong observation tool for many physical effects in optical lattices.

\begin{acknowledgments}
The work was supported by the Austrian Science Fund (FWF), through the Lise-Meitner Program (M977).
\end{acknowledgments}

\appendix

\section{Linear optical spectra using the input-output formalism}

Here we calculate the linear optical spectra by adopting the input-output formalisms. The input and output fields are given by
\begin{eqnarray}
H_U&=&\sum_{\bf k}\int d\omega_{\bf k}\ \hbar\omega_{\bf k}\ b^{\dagger}_{\bf k}(\omega_{\bf k})b_{\bf k}(\omega_{\bf k}), \nonumber \\
H_L&=&\sum_{\bf k}\int d\omega_{\bf k}\ \hbar\omega_{\bf k}\ c^{\dagger}_{\bf k}(\omega_{\bf k})c_{\bf k}(\omega_{\bf k}),
\end{eqnarray}
where $b^{\dagger}_{\bf k}(\omega_{\bf k}),\ b_{\bf k}(\omega_{\bf k})$ and $c^{\dagger}_{\bf k}(\omega_{\bf k}),\ c_{\bf k}(\omega_{\bf k})$ are the creation and annihilation operators of an external mode with in-plane wave vector ${\bf k}$ and frequency $\omega_{\bf k}$, above and below the cavity, respectively. The coupling between the external and internal fields at the cavity mirrors are given by
\begin{eqnarray}
V_U&=&\sum_{\bf k}\int d\omega_{\bf k}\ i\hbar u(\omega_{\bf k})\ \left[b^{\dagger}_{\bf k}(\omega_{\bf k})a_{\bf k}-a^{\dagger}_{\bf k}b_{\bf k}(\omega_{\bf k})\right], \nonumber \\
V_L&=&\sum_{\bf k}\int d\omega_{\bf k}\ i\hbar v(\omega_{\bf k})\ \left[c^{\dagger}_{\bf k}(\omega_{\bf k})a_{\bf k}-a^{\dagger}_{\bf k}c_{\bf k}(\omega_{\bf k})\right], \nonumber \\
\end{eqnarray}
where $u(\omega_{\bf k})$ and $v(\omega_{\bf k})$ are the coupling parameters at the upper and the lower mirrors. The coupling is between external and internal photons with the same in-plane wave vector. Here we consider light with a fixed polarization. For a fixed in-plane wave vector, we drop ${\bf k}$, to get the total Hamiltonian
\begin{eqnarray}
H&=&\sum_{r}\hbar\Omega_{r}\ A^{\dagger}_{r}A_{r} \nonumber \\
&+&\int d\omega\ \hbar\omega\ b^{\dagger}(\omega)b(\omega)+\int d\omega\ \hbar\omega\ c^{\dagger}(\omega)c(\omega) \nonumber \\
&+&\int d\omega\sum_r\ i\hbar u(\omega)\ \left[Y^{r\ast}\ b^{\dagger}(\omega)A_r-Y^{r}\ A_r^{\dagger}b(\omega)\right] \nonumber \\
&+&\int d\omega\sum_r\ i\hbar v(\omega)\ \left[Y^{r\ast}\ c^{\dagger}(\omega)A_r-Y^{r}\ A_r^{\dagger}c(\omega)\right], \nonumber \\
\end{eqnarray}
where, using the inverse transformation of Eqs.(\ref{TRANSFOR}), the cavity photon operator is written in terms of polariton operators $a=\sum_rY^{r\ast}A_r$. The external radiation field operator equations of motion are
\begin{eqnarray}
\frac{d}{dt}b(\omega)&=&-i\omega\ b(\omega)+u(\omega)\sum_rY^{r\ast}\ A_r,  \nonumber \\
\frac{d}{dt}c(\omega)&=&-i\omega\ c(\omega)+v(\omega)\sum_rY^{r\ast}\ A_r.
\end{eqnarray}
and the polariton operator equation of motion is
\begin{equation}
\frac{d}{dt}A_r=-i\Omega_r\ A_r-\int d\omega\ Y^r\left[u(\omega)\ b(\omega)+v(\omega)\ c(\omega)\right].
\end{equation}

The above system of equations are solved by using the input-output formalism \cite{Gardiner}. The external field operator equations are solved formally in using initial and final conditions, which are substituted back in the polariton operator equation and evaluated using a standard Markov approximation. For the upper mirror we define $\gamma_U=2\pi u^2(\omega)$, and for the lower one $\gamma_L=2\pi v^2(\omega)$, which are related to the mirror reflectivity as $\gamma_L\propto (1-R_L)$. Finally this gives:
\begin{eqnarray}
\frac{d}{dt}A_r&=&-i\Omega_r\ A_r+Y^r\left(\sqrt{\gamma_U}\ b_{in}+\sqrt{\gamma_L}\ c_{in}\right)-\gamma Y^r\ a, \nonumber \\
\frac{d}{dt}A_r&=&-i\Omega_r\ A_r-Y^r\left(\sqrt{\gamma_U}\ b_{out}+\sqrt{\gamma_L}\ c_{out}\right)+\gamma Y^r\ a, \nonumber \\
\end{eqnarray}
where $\gamma=(\gamma_U+\gamma_L)/2$, and we defined the input fields $a_{in},\ c_{in}$, and the output fields $a_{out},\ c_{out}$. The boundary conditions at the two mirrors are $\sqrt{\gamma_U}\ a=b_{in}+b_{out}$, and $\sqrt{\gamma_L}\ a=c_{in}+c_{out}$. Exciton damping is included by using a complex polariton eigenfrequency $\bar{\Omega}_r\equiv\Omega_r-i\Gamma_r$, where $\Gamma_r=\Gamma_{ex}|X^r|^2$, and where $\Gamma_{ex}$ is the effective exciton damping rate.

By taking the Fourier transform we solve these equations to get
\begin{eqnarray}
i(\bar{\Omega}_r-\omega)\ \tilde{A}_r&=&Y^r\left(\sqrt{\gamma_U}\ \tilde{b}_{in}+\sqrt{\gamma_L}\ \tilde{c}_{in}\right)-\gamma Y^r\ \tilde{a}, \nonumber \\
i(\bar{\Omega}_r-\omega)\ \tilde{A}_r&=&-Y^r\left(\sqrt{\gamma_U}\ \tilde{b}_{out}+\sqrt{\gamma_L}\ \tilde{c}_{out}\right)+\gamma Y^r\ \tilde{a}, \nonumber \\
\sqrt{\gamma_U}\ \tilde{a}&=&\tilde{b}_{in}+\tilde{b}_{out}\ ,\ \sqrt{\gamma_L}\ \tilde{a}=\tilde{c}_{in}+\tilde{c}_{out}.
\end{eqnarray}
For single side pumping, $\tilde{c}_{in}=0$, we then can directly express the output fields as a function of the input field, with average photon numbers defined by $n_{in}=\langle \tilde{b}_{in}^{\dagger}\tilde{b}_{in}\rangle$, $n_{out}^U=\langle \tilde{b}_{out}^{\dagger}\tilde{b}_{out}\rangle$, and $n_{out}^L=\langle \tilde{c}_{out}^{\dagger}\tilde{c}_{out}\rangle$. The transmission coefficient $T$ and the reflection coefficient $R$ are defined by $R(\omega)=n_{out}^U/n_{in}$, and $T(\omega)=n_{out}^L/n_{in}$. The absorption, $A$, is calculated from the relation $R(\omega)+T(\omega)+A(\omega)=1$. The results are given in Eqs.(\ref{SPECTRA}).

\end{document}